\documentclass[twocolumn,showpacs,prd,floatfix,axodraw]{revtex4}
\usepackage{mathrsfs}
\usepackage{graphicx,,booktabs,bm}
\usepackage{overpic}
\usepackage{color}
\usepackage{amssymb}

\usepackage{mathrsfs,bm,amsmath,amssymb}
\usepackage{longtable,lscape}
\usepackage{txfonts}
\usepackage{amssymb}
\usepackage{indentfirst}
\usepackage{graphicx,,booktabs}
\usepackage{multirow}
\usepackage{color}
\usepackage{amssymb}

\definecolor{cover}{rgb}{0.77,0.87,0.88}
\definecolor{blueone}{rgb}{0.1,0.1,.7}
\definecolor{citec}{rgb}{0.14,0.47,0.09}
\definecolor{two}{rgb}{0.0,0.5,0.}
\definecolor{three}{rgb}{.5,.1,0.15}
\usepackage[bookmarks=true,bookmarksopen=false,plainpages=false,breaklinks=true,
   bookmarksnumbered=true,hypertexnames=false,
   filecolor=blue,urlcolor=three,menucolor=three,
   linkcolor=three,citecolor=blueone, colorlinks,
   anchorcolor=blue,runcolor=pink,frenchlinks=red
   pdfstartview=FitH,pdftitle=title,%
   pdfauthor=author]{hyperref}

\begin{document}
\title{{{Strong decays of the $P_{cs}(4459)$ as a $\Xi_c\bar{D}^{*}$ molecule}}}

\author{Feng Yang$^1$}
\author{Yin Huang$^{1,3}$}
\author{Hong Qiang Zhu$^{2}$}
\email{20132013@cqnu.edu.cn}

\affiliation{$^1$School of Physical Science and Technology, Southwest Jiaotong University, Chengdu 610031,China\\
$^2$College of Physics and Electronic Engineering, Chongqing Normal University, Chongqing 401331,China\\
$^3$Asia Pacific Center for Theoretical Physics,
Pohang University of Science and Technology, Pohang 37673, Gyeongsangbuk-do,
South Korea}

\date{\today}
\begin{abstract}
In this work,  we study the strong decay of the newly observed $P_{cs}(4459)$, assuming that it is a pure $\Xi_c\bar{D}^{*}$
molecular state.  Considering two possible spin-parity assignments $J^P=1/2^{-}$ and $J^P=3/2^{-}$, the partial decay widths of
the $\Xi_c\bar{D}^{*}$ molecular state into $J/\psi{}\Lambda$, $D_s^{-}\Lambda_c^{+}$, and $\bar{D}\Xi_c^{(')}$ final states through
hadronic loops are evaluated with the help of the effective Lagrangians. In comparison with the LHCb data, the $S$-wave $\bar{D}^{*}\Xi_c$
molecular with $J^{P}=1/2^{-}$ assignment for $P_{cs}(4459)$ is supported by our study, while the $P_{cs}(4459)$ in spin-parity
$J^P=3/2^{-}$ case may be explained as an $S$-wave coupled bound state with lager $\Xi_c\bar{D}^{*}$ component.  In addition, the calculated
partial decay widths with $J^P=1/2^{-}$ $\Xi_c\bar{D}^{*}$ molecular state picture indicates that allowed decay mode, $\bar{D}\Xi_c^{'}$,
may have the biggest branching ratio. The experimental measurements for this strong decay process could be a crucial test for the molecule
interpretation of the $P_{cs}(4459)$.
\end{abstract}


\maketitle
\section{INTRODUCTION}
Thanks to the great progress of the experiment in the past several decades, many baryons that cannot be ascribed into
3-quark(qqq) configurations have been reported~\cite{Zyla:2020zbs}.  In particular, three narrow hidden-charm pentaquarks, namely
$P_c(4312)$,$P_c(4440)$, and $P_c(4450)$, were observed by the LHCb Collaboration in the $J/\psi{}p$ invariant mass distributions of
the $\Lambda_b\to{J/\psi}pK$ decay~\cite{Aaij:2019vzc}.  Very recently, a new possible strange hidden charm pentaquark $P_{cs}(4459)$
was reported by the LHCb Collaboration in the $J/\psi\Lambda$ final state from the $\Xi_b^{-}\to{}J/\psi{}\Lambda{}K^{-}$
process~\cite{Aaij:2020gdg}.  The observed resonance masses and widths are
\begin{align}
M&=4458.8\pm{}2.9^{+4.7}_{-1.1}~~~{\rm MeV},\nonumber\\
\Gamma&=17.3\pm{}6.5^{+8.0}_{-5.7}~~~~~ {\rm MeV},
\end{align}
respectively.  From the observed decay mode, the isospin of this state is zero.  Although the spin and parity
remain undetermined, it is very helpful to understand the spectroscopy of the hidden-charm pentaquark.

Before the discovery of the $P_{cs}(4459)$ state, there were already a few theoretical studies on the existence of the
strange hidden-charm
pentaquarks~\cite{Wu:2010vk,Wang:2019nvm,Anisovich:2015zqa,Wang:2015wsa,Feijoo:2015kts,Lu:2016roh,Chen:2016ryt,Xiao:2019gjd,Chen:2015sxa}.
And Refs.~\cite{Lu:2016roh,Feijoo:2015kts,Chen:2015sxa} proposed to search for these strange hidden-charm pentaquarks states
in $\Xi_b(\Lambda_b)\to{}J/\psi\Lambda{}\eta(K)$.  In particular, Wang et al.found two isoscalar $\Xi_c\bar{D}^{*}$
molecular states with the chiral effective field theory~\cite{Wang:2019nvm}; the masses were predicted considering the
$\Xi_c\bar{D}^{*}$ molecules for $J^P=1/2^{-}$and $J^P=3/2^{-}$ as 4456.9 and 4463.0 MeV, respectively, which can
be associated with the experimental $P_{cs}(4459)$.

The observation of the $P_{cs}(4459)$ inspires a large number of theoretical studies about its internal structure.
The QCD sum rule suggests that the newly observed state $P_{cs}(4459)$ is most likely to be considered a
scalar-diquark-scalar-diquark-antiquark with the spin parity $J^P=1/2^{-}$~\cite{Wang:2020eep}. Based on the analysis of the
two-body allowed strong decays $P_{cs}(4459)\to{}J/\psi{}\Lambda$ with a similar method above but considering different details,
the same conclusion was drawn from Ref.~\cite{Azizi:2021utt} that $P_{cs}(4459)$ can be understood as a compact pentaquark
nature of diquark-diquark-antiquark form.

Because the mass is just 19 MeV below the $\Xi_c\bar{D}^{*}$ threshold, the recent observation of $P_{cs}(4459)$ also favors the
molecular state interpretations of the hidden-charm pentaquarks.  Indeed, the $\Xi_c\bar{D}^{*}$ molecular
explanation for the $P_{cs}(4459)$ was discussed using effective field formalism, and the spin of the $P_{cs}(4459)$ state was
suggested to be $J=3/2$~\citep{Peng:2020hql}.  The molecular explanation for the $P_{cs}(4459)$ was also studied in Ref.~\cite{Chen:2020uif}
using QCD sum rule method, and the results supported its possibility to be a $\Xi_c\bar{D}^{*}$ molecular state with either
$J^P=1/2^{-}$ or $J^P=3/2^{-}$.  By analyzing the spectroscopy using the one-boson-exchange model, Refs.~\cite{Chen:2020kco,Zhu:2021lhd}
assigned the $P_{cs}(4459)$ as a coupled bound states with spin-parity $J^P=3/2^{-}$ instead of pure $\Xi_c\bar{D}^{*}$ molecular.
A possible explanation for these results of Refs.~\cite{Peng:2020hql,Chen:2020uif,Chen:2020kco,Zhu:2021lhd} is that the $P_{cs}(4459)$
has a predominant $\Xi_c\bar{D}^{*}$ component and that the lager $\Xi_c\bar{D}^{*}$ component makes the interaction between a $\bar{D}^{*}$ meson and a
$\Xi_c$ baryon strong enough to form a bound state with a mass roughly 4459 MeV~\cite{Zhu:2021lhd}.  This make it possible to search for the
$P_{cs}(4459)$ in the $\Xi_b^{-}\to{}J/\psi\Lambda{}K^{-}$ by assuming the $P_{cs}(4459)$ is a molecular mainly composed of
$\bar{D}^{*}\Xi_c$~\cite{Lu:2021irg}.

Until now, the nature of the observed $P_{cs}(4459)$ baryon remains unclear.  In particular, the quantum numbers of the $P_{cs}(4459)$
was not determined by the experiment, and from different studies, there are different assumptions for its quantum numbers
and substructure.   In this work, we will calculate the strong decay pattern of S-wave $\Xi_c\bar{D}^{*}$ molecular state within the
effective Lagrangians approach, and find the relation between the $\bar{D}^{*}\Xi_c$ molecular state and the $P_{cs}(4459)$ state
by comparing with the LHCb observation.  This idea is inspired by the allowed strong decay that will almost saturate its total
decay width.

This paper is organized as follows. The theoretical
formalism is explained in Sec.~\ref{Sec: formulism}. The predicted partial
decay widths are presented in Sec.~\ref{Sec: results}. The last section includes the discussion
and our summary.

\section{THEORETICAL FORMALISM}\label{Sec: formulism}
\begin{figure}[h!]
\begin{center}
\includegraphics[bb=80 440 750 710, clip, scale=0.55]{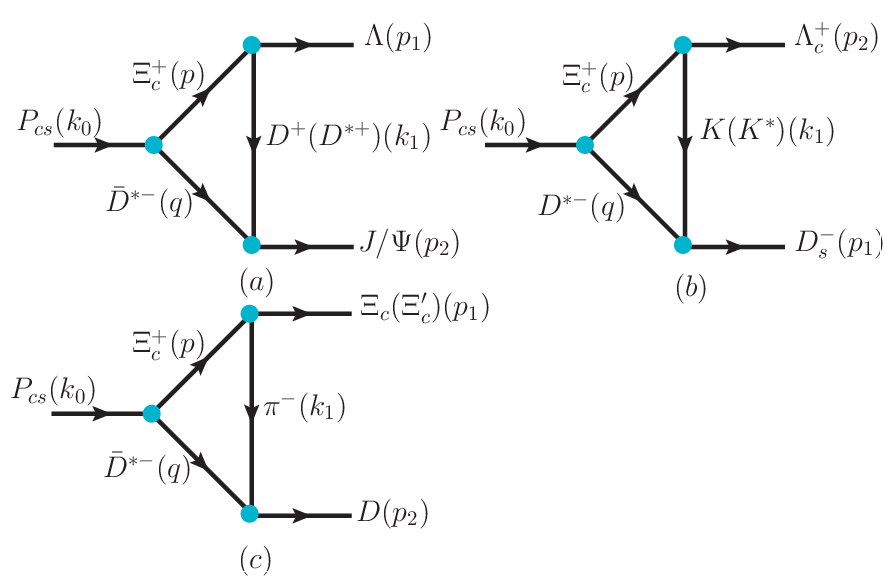}
\caption{Feynman diagrams for the $P_{cs}\to{}J/\psi\Lambda,\bar{D}\Xi_c^{(')}$, and $D_s^{-}\Lambda_c^{+}$  decay processe.
The contributions from the t-channel $D$ and $D^{*}$ exchange(a),$K$ and $K^{*}$ exchange(b), and $\pi$ exchange (c) are
considered.  We also show the definition of the kinematical $(k_0,p,q,p_1, p_2,k_1)$ that we use in the present calculation.}
\label{cc1}
\end{center}
\end{figure}

In this work, we study the strong decay widths $P_{cs}\to{}J/\psi\Lambda,D\Xi_c^{(')}$, and $D_s^{-}\Lambda_c^{+}$
in the molecular scenario with different spin-parity assignments for the $P_{cs}$.   The Feynman diagrams for the hadronic
decay of the $\Xi_c\bar{D}^{*}$ molecular state into $J/\psi\Lambda,\bar{D}\Xi_c^{(')}$ and $D_s^{-}\Lambda_c^{+}$ mediated by the
exchange of the $\pi$, $K^{(*)}$, and $D^{(*)}$ mesons are shown in Fig.~\ref{cc1}.

To compute the diagrams shown in Fig.~\ref{cc1}, we need the effective Lagrangian densities for the relevant interaction
vertices. As mentioned above, the $P_{cs}(4459)$ resonance is identified as an $S$-wave $\Xi_c\bar{D}^{*}$ molecule.  The molecular
structure of the $P_{cs}(4459)$ baryon with $J^P=1/2^{-}$ is described by the following Lagrangian~\cite{Xiao:2019mvs}
\begin{align}
\mathcal{L}_{P_{c s} \Xi_{c} \bar{D}^{*}}&=g_{p_{c s}} \bar{P}_{c s}(x) \gamma^{\mu}\gamma^{5}\nonumber\\
	                                         &\times\int dy\phi(y^2)\sum{}C_i\Xi_{c}(x+\omega_{\bar{D}^{*}}y)\bar{D}^{*} _{\mu}(x-\omega_{\Xi_c}y)\label{eqw1},
\end{align}
while for $J^P=3/2^{-}$ the Lagrangian contains a derivative $\bar{D}^{*}\Xi_c$ coupling~\cite{Xiao:2019mvs}
\begin{align}
\mathcal{L}_{P_{c s} \Xi_{c} \bar{D}^{*}}&=-ig_{P_{c s}} \bar{P}_{c s}^{\mu}(x) \nonumber\\
	                                         &\times\int dy\phi(y^2)\sum{}C_i\Xi_{c}(x+\omega_{\bar{D}^{*}}y)\bar{D}^{*} _{\mu}(x-\omega_{\Xi_c}y)\label{eqw2},
\end{align}
where $\omega_{i}=m_{i}/(m_{i}+m_{j})$.  In the above Lagrangians, the effective correlation function $\Phi(y^2)$ shows the distribution of the components
in the hadronic molecule $P_{cs}(4459)$ state.  Moreover, the role of the correlation function $\Phi(y^2)$ is also to avoid the Feynman diagram's
ultraviolet divergence, and his Fourier transform should vanish quickly in the ultraviolet region in the Euclidean space.  We adopt the form as
used in Refs.~\cite{Dong:2008gb,Dong:2017rmg}
\begin{align}
\Phi(-p^2)\doteq{}\exp(-p_E^2/\Lambda^2),
\end{align}
where $p_E$ is the Euclidean Jacobi momentum. At present, the value of $\Lambda$ still could not be accurately determined from first principles,
 and therefore, it better determined by experimental data. The experimental total widths of some
states~\cite{Chen:2016qju,Xiao:2016mho,Cleven:2013mka,Dong:2008gb,Dong:2017rmg,Branz:2009yt,Faessler:2007gv} that can be considered as molecules
can be well explained with $\Lambda$ = 1.0 GeV.  We therefore assume $\Lambda$ = 1.0 GeV in this work to study whether the $P_{cs}(4459)$ can be
interpreted as a molecule composed of $\bar{D}^{*}\Xi_c$.

The coupling constants $g_{P_{cs}}$ in Eq.~\ref{eqw1} and Eq.~\ref{eqw2} can be calculated by employing the compositeness
condition~\cite{Weinberg:1962hj,Salam:1962ap,Dong:2009tg,Oller:2017alp,Guo:2015daa}.  This condition requires that the renormalization constant of the
hadronic molecular wave function is equal to zero
\begin{align}
&1-\frac{d\Sigma_{P_{cs}}}{dk_0}\mid_{k_0=m_{P_{cs}}}=0,    ~~~~~~~~~~j=\frac{1}{2}\\
&1-\frac{d\Sigma_{P_{cs}}^T}{dk_0}\mid_{k_0=m_{P_{cs}}}=0,  ~~~~~~~~~~j=\frac{3}{2}
\end{align}
where $k_0^2=m^2_{P_{cs}}$ with $k_0$, $m_{P_{cs}}$ denotes the four momenta and the mass of the $P_{cs}(4459)$, respectively,
Here, we set $m_{P_{cs}}=m_{\Xi_c}+m_{\bar{D}^{*}}-E_b$ with $m_{\Xi_c}$, $m_{\bar{D}^{*}}$, with $E_b$ being the mass of the
components $\Xi_c$, $\bar{D}^{*}$, and the binding energy of $P_{cs}(4459)$, respectively.  The $\Sigma_{P_{cs}}$ is the self-energy
of the hadronic molecule $P_{cs}(4459)$, and the $\Sigma^{T}_{P_{cs}}$ is the transverse part of the self-energy operator
$\Sigma^{\mu\nu}_{P_{cs}},$ related to $\Sigma^{\mu\nu}_{P_{cs}}$ via
\begin{align}
\Sigma^{\mu\nu}_{P_{cs}}=(g_{\mu\nu}-\frac{k_0^{\mu}k_0^{\nu}}{k_0^2})\Sigma^{T}_{P_{cs}}+\cdots.
\end{align}

\begin{figure}[htbp]
\begin{center}
\includegraphics[bb=140 620 750 720, clip,scale=0.75]{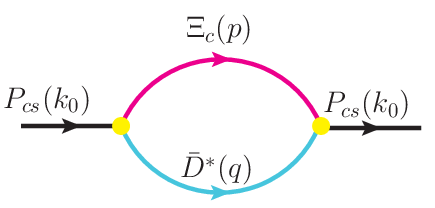}
\caption{Self-energy of the $P_{cs}(4459)$ state.}
\label{t-mass}
\end{center}
\end{figure}

The Feynman diagram describing the self-energy of the $P_{cs}(4459)$ state is presented in Fig.~\ref{t-mass}.
With the effective Lagrangians in Eq.~\ref{eqw1} and Eq.~\ref{eqw2}, we can compute the Feynman diagrams shown
in Fig.~\ref{t-mass}, and obtain the self-energy of the $P_{cs}(4459)$,
\begin{align}
\Sigma_{P_{cs}}^{1/2}&=\sum_{i}C_ig_{P_{cs}}^2\int \frac{d^4q}{(2\pi)^4} \Phi^2[(p\omega_{\bar{D}^{*}}-q\omega_{\Xi_c})^2]\gamma^{\mu}\gamma^{5}\nonumber\\
&\times \frac{i(p\!\!\!/+m_{\Xi_{c}})}{p^2-m_{\Xi_{c}}^2}\gamma^{\nu}\gamma^{5}\frac{i(-g^{\mu\nu}+q^{\mu}q^{\nu}/m_{\bar{D}^{*}}^2)}{q^2-m_{\bar{D}^{*}}^2},\\
\Sigma_{P_{cs}}^{3/2}&=\sum_iC_ig_{P_{cs}}^2\int \frac{d^4q}{(2\pi)^4} \Phi^2[(p\omega_{\bar{D}^{*}}-q\omega_{\Xi_c})^2]\nonumber\\
&\times\frac{i(p\!\!\!/+m_{\Xi_{c}})}{p^2-m_{\Xi_{c}}^2}\frac{i(-g^{\mu\nu}+q^{\mu}q^{\nu}/m_{\bar{D}^{*}}^2)}{q^2-m_{\bar{D}^{*}}^2},
\end{align}
where isospin symmetry implies that
\begin{align}
\begin{split}
C_{i}= \left \{
\begin{array}{ll}
    -1/\sqrt{2},                           & i=\Xi_c^0\bar{D}^{*0}\\
    ~~~~~~~~~~~~~~~~~~~                    & ~~~~~~~~~~~~~~~~~~~~\\
    ~1/\sqrt{2},                           & i=\Xi_c^{+}D^{*-}
\end{array}
\right.
\end{split}
\end{align}
with the following isospin assignments for the $\Xi_{c}$ and $\bar{D}^*$
\begin{align}
\left(
\begin{array}{c}
 \Xi_c^+ \\ \Xi_c^0
\end{array}
\right)
\sim
\left(
\begin{array}{c}
|\frac{1}{2}, \frac{1}{2} \rangle \\	|\frac{1}{2}, -\frac{1}{2} \rangle
\end{array}
\right)	,
\left(
\begin{array}{c}
D^{*-} \\ \bar{D}^{*0}
\end{array}
\right)
\sim
\left(
\begin{array}{c}
-|\frac{1}{2},- \frac{1}{2} \rangle \\	|\frac{1}{2}, \frac{1}{2} \rangle
\end{array}
\right).	
\end{align}

To compute the amplitudes of the diagrams shown in Fig.~\ref{cc1}, the effective Lagrangian densities related
to the final state are required, which are~\cite{Chen:2021tip}
\begin{align}
	{\cal{L}}_{PPV}&=\frac{iG}{2\sqrt{2}}\langle\partial^{\mu}P(PV_{\mu}-V_{\mu}P)\rangle,\\
		{\cal{L}}_{VVP}&=\frac{G'}{\sqrt{2}}\epsilon^{\mu\nu\alpha\beta}\langle\partial_{\mu}V_{\nu}\partial_{\alpha}V_{\beta}P\rangle,\\
		{\cal{L}}_{VVV}&=\frac{iG}{2\sqrt{2}}\langle \partial^{\mu}V^{\nu}(V_{\mu}V_{\nu}-V_{\nu}V_{\mu})\rangle.
\end{align}
where the coupling constants are $G=12.00$ and $G'=55.51$.   The symbol $V_{\mu}$ and $P$ represents the vector and pseudoscalar fields of
the 16-plet of the $\rho$ and $\pi$ meson,respectively,
\begin{align}
	P&=\left(
	\begin{array}{cccc}
		\frac{\pi^{0}}{\sqrt{2}}+\frac{\eta}{\sqrt{6}}+\frac{\eta_{c}}{\sqrt{12}} &   \pi^{+}   &  K^{+}  & \bar{D}^{0} \\
		\pi^{-} & -\frac{\pi^{0}}{\sqrt{2}}+\frac{\eta}{\sqrt{6}}+\frac{\eta_{c}}{\sqrt{12}} & K^{0} & D^{-}\\
		K^{-} & \bar{K}^{0} & -\frac{2\eta}{\sqrt{6}}+\frac{\eta_{c}}{\sqrt{12}} & D_{s}^{-}\\
		D^{0} & D^{+} & D_{s}^{+} & -\frac{3\eta_{c}}{\sqrt{12}}
	\end{array}
	\right) ,\nonumber\\
V&=\left(
\begin{array}{cccc}
	\frac{\rho^{0}}{\sqrt{2}}+\frac{\omega_8}{\sqrt{6}}+\frac{J/\Psi}{\sqrt{12}} &   \rho^{+}   &  K^{*+}  & \bar{D}^{*0} \\
	\rho^{-} & -\frac{\rho^{0}}{\sqrt{2}}+\frac{\omega_8}{\sqrt{6}}+\frac{J/\Psi}{\sqrt{12}} & K^{*0} & D^{*-}\\
	K^{*-} & \bar{K}^{*0} & -\frac{2\omega_{8}}{\sqrt{6}}+\frac{J/\Psi}{\sqrt{12}} & D_{s}^{*-}\\
	D^{*0} & D^{*+} & D_{s}^{*+} & -\frac{3J/\Psi}{\sqrt{12}}
\end{array}
\right)
\end{align}
with $\omega_{8}=\omega cos\theta +\phi sin\theta$ and $sin\theta=-0.761$.

The SU(4) invariant interaction Lagrangians between baryons and pseudoscalar mesons as well as between baryons and vector mesons
can be written, respectively, as~\cite{Liu:2001ce}
\begin{align}
{\cal{L}}_{PBB}&=ig_{p}(a\phi^{*\alpha\mu\nu}\gamma_{5}P_{\alpha}^{\beta}\phi_{\beta\mu\nu}+b\phi^{*\alpha\mu\nu}\gamma_{5}P^{\beta}_{\alpha}\phi_{\beta\nu\mu}),\\	
{\cal{L}}_{VBB}&=ig_{v}(c\phi^{*\alpha\mu\nu}\gamma V_{\alpha}^{\beta}\phi_{\beta\mu\nu}+d\phi^{*\alpha\mu\nu}\gamma  V^{\beta}_{\alpha}\phi_{\beta\nu\mu}),
\end{align}
where $g_p$ and $g_v$ are universal baryon-pseudoscalar-meson and baryon-vector-meson coupling constants, and a,b,c, and d are constants.
In the SU(4) quark model, baryons belong to the 20-plet states~\cite{Liu:2001ce}.  These states can be conveniently expressed by
tensors $\phi_{\mu\nu\lambda}$, where $\mu,\nu, $ and $\lambda$ run from 1 to 4, and satisfy the conditions
\begin{align}
	\phi_{\mu\nu\lambda}+\phi_{\nu\lambda\mu}+\phi_{\lambda\mu\nu}=0,~~\phi_{\mu\nu\lambda}=\phi_{\nu\mu\lambda},
\end{align}
with
\begin{align}
	&p=\phi_{112}, ~~ n=\phi_{221}, ~~\Lambda=\sqrt{2/3}(\phi_{321}-\phi_{312}),\nonumber\\
	&\Sigma^{+}=\phi_{113},~~  \Sigma^{0}=\sqrt{2}\phi_{123},~~\Sigma^{-}=\phi_{223},\nonumber\\
	&\Xi^{0}=\phi_{331},~~\Xi^{-}=\phi_{332},~~\Sigma_{c}^{++}=\phi_{114},\nonumber\\
	&\Sigma_{c}^{+}=\phi_{124},~~\Sigma_{c}^{0}=\phi_{224},~~\Xi_{c}^{+}=\phi_{134},\nonumber
\end{align}
\begin{align}
	&\Xi_{c}^{0}=\phi_{234},~~\Xi_{c}^{+'}=\sqrt{2/3}(\phi_{413}-\phi_{431}),~~\Xi_{c}^{0'}=\sqrt{2/3}(\phi_{423}-\phi_{432}),\nonumber\\
	&\Lambda_{c}^{+}=\sqrt{2/3}(\phi_{421}-\phi_{412}),~~\Omega_{c}^{0}=\phi_{334},~~\Xi_{cc}^{++}=\phi_{441},\nonumber\\
	&\Xi_{cc}^{+}=\phi_{442},~~\Omega_{cc}^{+}=\phi_{443}.
\end{align}
Writing explicitly, we obtain the following interaction Lagrangians,
\begin{align}
	{\cal{L}}_{PBB}&=ig_{p}\bigg[\frac{1}{\sqrt{2}}\bigg( a-\frac{5}{4}b \bigg)\bar{N}\gamma_{5}\vec{\tau}\cdot \vec{\pi}N \nonumber\\
	               &+ \frac{3\sqrt{6}}{8}(b-a)(\bar{N}\gamma_{5}K\Lambda+\bar{N}\gamma_{5}\bar{D}\Lambda_{c})+\cdots \bigg],\\
	{\cal{L}}_{VBB}&=ig_{v}\bigg[\frac{1}{\sqrt{2}}\bigg( c-\frac{5}{4}d \bigg)\bar{N}\gamma_{\mu}\rho^{\mu}N \nonumber\\
	               &+ \frac{3\sqrt{6}}{8}(d-c)(\bar{N}\gamma_{\mu}K^{*\mu}\Lambda+\bar{N}\gamma_{\mu}\bar{D}^{\mu}\Lambda_{c})+\cdots \bigg],
\end{align}
where we know $b/a=5.3,d/c=0.5,g_{\pi NN}=13.5,g_{\rho NN}=3.25$.

Putting all the pieces together, we obtain the following amplitudes
\begin{align}
	{\cal{M}}^{1/2^{-}}_{a}&=\bar{\mu}(p_1)\bigg(\frac{1}{4\sqrt{3}}g_{\Xi_{c}^+\Lambda D^+}g_{D^{*-}D^{+}J/\Psi}g_{p_{c s}}\int\frac{d^{4}k_{1}}{(2\pi)^{4}}
	\nonumber\\
	&\times\Phi[(p\omega_{\bar{D}^{*-}}-q\omega_{\Xi_c^+})^2]\epsilon_{\mu\nu\alpha\beta}\gamma^5\frac{1}{k_{1}^{2}-m_{D^{+}}^2}\nonumber\\
	&\times(p_{2}^{\mu}q^{\alpha}g^{\nu\theta}g^{\beta\eta}-3p_{2}^{\alpha}q^{\mu}g^{\beta\theta}g^{\nu\eta})\frac{-g^{\sigma\eta}+q^{\sigma}q^{\eta}/m_{D^{*-}}^2}
	{q^2-m_{D^{*-}}^2}\gamma^{\sigma}\gamma^5\nonumber\\
	&\times\frac{p\!\!\!/+m_{\Xi_{c}^{+}}}{p^2-m_{\Xi_{c}^{+}}^2}-i\frac{\sqrt{3}}{3}g_{\Xi_{c}^+\Lambda D^{*+}}g_{D^{*-}D^{*+}J/\Psi}g_{p_{c s}}\int\frac{d^{4}k_{1}}{(2\pi)^{4}}\nonumber\\
	&\times\Phi[(p\omega_{\bar{D}^{*-}}-q\omega_{\Xi_c^+})^2]\gamma_{\lambda}\frac{-g^{\lambda\rho}+k_{1}^{\lambda}k_{1}^{\rho}/m_{D^{*+}}^2}{k_{1}^2-m_{D^{*+}}^2}\nonumber\\
	&\times(k_{1}^{\theta}g^{\eta\rho}-p_{2}^{\eta}g^{\rho\theta}+q^{\rho}g^{\theta\eta})\frac{-g^{\sigma\eta}+q^{\sigma}q^{\eta}/m_{D^{*-}}^2}{q^2-m_{D^{*-}}^2}\nonumber\\
	&\times\gamma^{\sigma}\gamma^5\frac{p\!\!\!/+m_{\Xi_{c}^{+}}}{p^2-m_{\Xi_{c}^{+}}^2} \bigg)\varepsilon^{\theta}(p_2)\mu(k_0),\\
{\cal{M}}^{1/2^{-}}_{b}&=\bar{\mu}(p_1)\bigg(-\frac{1}{2}g_{P_{cs}}^{1/2}g_{\Lambda_{c}^{+}\Xi_{c}^{+}K}g_{D_{s}^{-}D^{*-}K}\int\frac{d^{4}k_{1}}{(2\pi)^{4}}\nonumber
\end{align}
\begin{align}
&\times\Phi[(p\omega_{\bar{D}^{*-}}-q\omega_{\Xi_c^+})]^2\gamma^{5}\frac{1}{k_{1}^{2}-m_{K}^{2}}\nonumber\\
&\times(k_{1}^{\eta}-p_{2}^{\eta})\frac{-g^{\sigma\eta}+q^{\sigma}q^{\eta}/m_{D^{*-}}^2}{q^2-m_{D^{*-}}^2}\gamma^{\sigma}
\gamma^5\frac{p\!\!\!/+m_{\Xi_{c}^{+}}}{p^2-m_{\Xi_{c}^{+}}^2}\nonumber\\
&-\frac{1}{2}ig_{P_{cs}}^{1/2}g_{\Lambda_{c}^{+}\Xi_{c}^{+}K^{*}}g_{D_{s}^{-}D^{*-}K^{*}}\int\frac{d^{4}k_{1}}{(2\pi)^{4}}\nonumber\\
&\times\Phi[(p\omega_{\bar{D}^{*-}}-q\omega_{\Xi_c^+})^2]\epsilon_{\mu\nu\alpha\beta}\gamma_{\lambda}\frac{-g^{\lambda\rho}+k_{1}^{\lambda}k_{1}^{\rho}/m_{K^{*}}^2}{k_{1}^2-m_{K^{*}}^2}\nonumber\\
&\times q^{\mu}k_{1}^{\alpha}g^{\nu\eta}g^{\beta\rho}\frac{-g^{\sigma\eta}+q^{\sigma}q^{\eta}/m_{D^{*-}}^2}{q^2-m_{D^{*-}}^2}\gamma^{\sigma}\nonumber\\&\times\gamma^5\frac{p\!\!\!/+m_{\Xi_{c}^{+}}}{p^2-m_{\Xi_{c}^{+}}^2} \bigg)\mu(k_0),\\
{\cal{M}}^{1/2^{-}}_{c}&=\bar{\mu}(p_1)\bigg(-\frac{1}{2}g_{P_{cs}^{1/2}}g_{\Xi_{c}^{+}\Xi_{c}^{0}\pi^{-}}g_{D^{*-}\pi^{-}D}\int\frac{d^{4}k_{1}}{(2\pi)^{4}}\nonumber\\
&\times\Phi[(p\omega_{\bar{D}^{*-}}-q\omega_{\Xi_c^+})^2]\gamma^{5}\frac{1}{k_{1}^{2}-m_{\pi^{-}}^{2}}(p_{2}^{\eta}-k_{1}^{\eta})\nonumber\\
	&\times\frac{-g^{\sigma\eta}+q^{\sigma}q^{\eta}/m_{D^{*-}}^2}{q^2-m_{D^{*-}}^2}\gamma^{\sigma}\gamma^5\frac{p\!\!\!/+m_{\Xi_{c}^{+}}}{p^2-m_{\Xi_{c}^{+}}^2} \bigg)\mu(k_0), \\
{\cal{M}}^{1/2^{-}}_{c'}&={\cal{M}}_{c}(g_{\Xi_{c}^{+}\Xi_{c}^{0}\pi^{-}}\rightarrow g_{\Xi_{c}^{+}\Xi_{c}^{'0}\pi^{-}},m_{\Xi_{c}^{0}}\rightarrow m_{\Xi_{c}^{'0}}),
\end{align}
and
\begin{align}
{\cal{M}}_{a}^{3/2^{-}}&=\bar{\mu}(p_1)\bigg(\frac{1}{4\sqrt{3}}ig_{\Xi_{c}^+\Lambda D^+}g_{D^{*-}D^{+}J/\Psi}g_{p_{c s}}\int\frac{d^{4}k_{1}}{(2\pi)^{4}}\nonumber\\
&\times\Phi[(p\omega_{\bar{D}^{*-}}-q\omega_{\Xi_c^+})^2]\epsilon_{\mu\nu\alpha\beta}\gamma^5\frac{1}{k_{1}^{2}-m_{D^{+}}^2}\nonumber\\
&\times(p_{2}^{\mu}q^{\alpha}g^{\nu\theta}g^{\beta\eta}-3p_{2}^{\alpha}q^{\mu}g^{\beta\theta}g^{\nu\eta})\frac{-g^{\sigma\eta}+q^{\sigma}q^{\eta}/m_{D^{*-}}^2}{q^2-m_{D^{*-}}^2}\nonumber\\
&\times\frac{p\!\!\!/+m_{\Xi_{c}^{+}}}{p^2-m_{\Xi_{c}^{+}}^2}+\frac{\sqrt{3}}{3}g_{\Xi_{c}^+\Lambda D^{*+}}g_{D^{*-}D^{*+}J/\Psi}g_{p_{c s}}\int\frac{d^{4}k_{1}}{(2\pi)^{4}}\nonumber\\
&\times\Phi[(p\omega_{\bar{D}^{*-}}-q\omega_{\Xi_c^+})^2]\gamma_{\lambda}\frac{-g^{\lambda\rho}+k_{1}^{\lambda}k_{1}^{\rho}/m_{D^{*+}}^2}{k_{1}^2-m_{D^*+}^2}\nonumber\\
&\times(k_{1}^{\theta}g^{\eta\rho}-p_{2}^{\eta}g^{\rho\theta}+q^{\rho}g^{\theta\eta})\frac{-g^{\sigma\eta}+q^{\sigma}q^{\eta}/m_{D^{*-}}^2}{q^2-m_{D^{*-}}^2}\
\nonumber\\
&\times\frac{p\!\!\!/+m_{\Xi_{c}^{+}}}{p^2-m_{\Xi_{c}^{+}}^2} \bigg)\varepsilon^{\theta}(p_2)\mu^{\sigma}(k_0),\\
{\cal{M}}^{3/2^{-}}_{b}&=\bar{\mu}(p_1)\bigg(-\frac{1}{2}ig_{P_{cs}}g_{\Lambda_{c}^{+}\Xi_{c}^{+}K}g_{D_{s}^{-}D^{*-}K}\int\frac{d^{4}k_{1}}{(2\pi)^{4}}\nonumber\\
&\times\Phi[(p\omega_{\bar{D}^{*-}}-q\omega_{\Xi_c^+})^2]\gamma^{5}\frac{1}{k_{1}^{2}-m_{K}^{2}}\frac{-g^{\sigma\eta}+q^{\sigma}q^{\eta}/m_{D^{*-}}^2}{q^2-m_{D^{*-}}^2}\nonumber\\
&\times\frac{p\!\!\!/+m_{\Xi_{c}^{+}}}{p^2-m_{\Xi_{c}^{+}}^2}+\frac{1}{2}g_{P_{cs}}g_{\Lambda_{c}^{+}\Xi_{c}^{+}K^{*}}g_{D_{s}^{-}D^{*-}K^{*}}\int\frac{d^{4}k_{1}}{(2\pi)^{4}}\nonumber\\
&\times\Phi[(p\omega_{\bar{D}^{*-}}-q\omega_{\Xi_c^+})^2]\epsilon_{\mu\nu\alpha\beta}\gamma_{\lambda}\frac{-g^{\lambda\rho}+k_{1}^{\lambda}k_{1}^{\rho}/m_{K^{*}}^2}{k_{1}^2-m_{K^{*}}^2}\nonumber\\
&\times\frac{-g^{\sigma\eta}+q^{\sigma}q^{\eta}/m_{D^{*-}}^2}{q^2-m_{D^{*-}}^2}\frac{p\!\!\!/+m_{\Xi_{c}^{+}}}{p^2-m_{\Xi_{c}^{+}}^2} \bigg)\mu^{\sigma}(k_0),
\end{align}
\begin{align}
{\cal{M}}^{3/2^{-}}_{c}&=\bar{\mu}(p_1)\bigg(\frac{1}{2}ig_{P_{cs}}g_{\Xi_{c}^{+}\Xi_{c}^{0}\pi^{-}}g_{D^{*-}\pi^{-}D}\int\frac{d^{4}k_{1}}{(2\pi)^{4}}\nonumber\\
&\times\Phi[(p\omega_{\bar{D}^{*-}}-q\omega_{\Xi_c^+})^2]\gamma^{5}\frac{1}{k_{1}^{2}-m_{\pi^{-}}^{2}}(p_{2}^{\eta}-k_{1}^{\eta})\nonumber\\
&\times\frac{-g^{\sigma\eta}+q^{\sigma}q^{\eta}/m_{D^{*-}}^2}{q^2-m_{D^{*-}}^2}\frac{p\!\!\!/+m_{\Xi_{c}^{+}}}{p^2-m_{\Xi_{c}^{+}}^2} \bigg)\mu^{\sigma}(k_0) \\%c
{\cal{M}}^{3/2^{-}}_{c'}&={\cal{M}}_{c}(g_{\Xi_{c}^{+}\Xi_{c}^{0}\pi^{-}}\rightarrow g_{\Xi_{c}^{+}\Xi_{c}^{'0}\pi^{-}},m_{\Xi_{c}^{0}}\rightarrow m_{\Xi_{c}^{'0}}).
\end{align}

Once the amplitudes are determined, the corresponding partial decay widths can be obtained, which read,
\begin{align}
\Gamma(P_{cs}\to)=\int\frac{1}{2J+1}\frac{1}{32\pi^2}\frac{|\vec{p}_1|}{m^2_{P_{cs}}}|{\cal{M}}|^2d\Omega
\end{align}
where the $J$ is the total angular momentum of the $P_{cs}(4459)$, $|\vec{p}_1|$ is the three-momenta of the decay
products in the center of mass frame, the overline indicates the sum over the polarization vectors of the final hadrons.
The $\Omega$ is the angle of the final particle in the rest frame of $P_{cs}(4459)$.

\section{RESULTS}\label{Sec: results}
In this work, we study the strong decays of the $P_{cs}(4459)$ to the two-body final states $J/\psi{}\Lambda$, $D_s^{-}\Lambda_c^{+}$,
and $\bar{D}\Xi_c^{(')}$, assuming that $P_{cs}(4459)$ is a $\bar{D}^{*}\Xi_c$ molecular state.  In order to obtain the two body
decay width through the triangle diagrams shown in Fig.\ref{cc1}, we first need to compute the coupling constants $g_{P_{cs}}$ relevant to
the effective Lagrangians listed in Eqs.~\ref{eqw1} and \ref{eqw2}.

\begin{figure}[htbp]
\begin{center}
\includegraphics[bb=50 10 750 520, clip,scale=0.30]{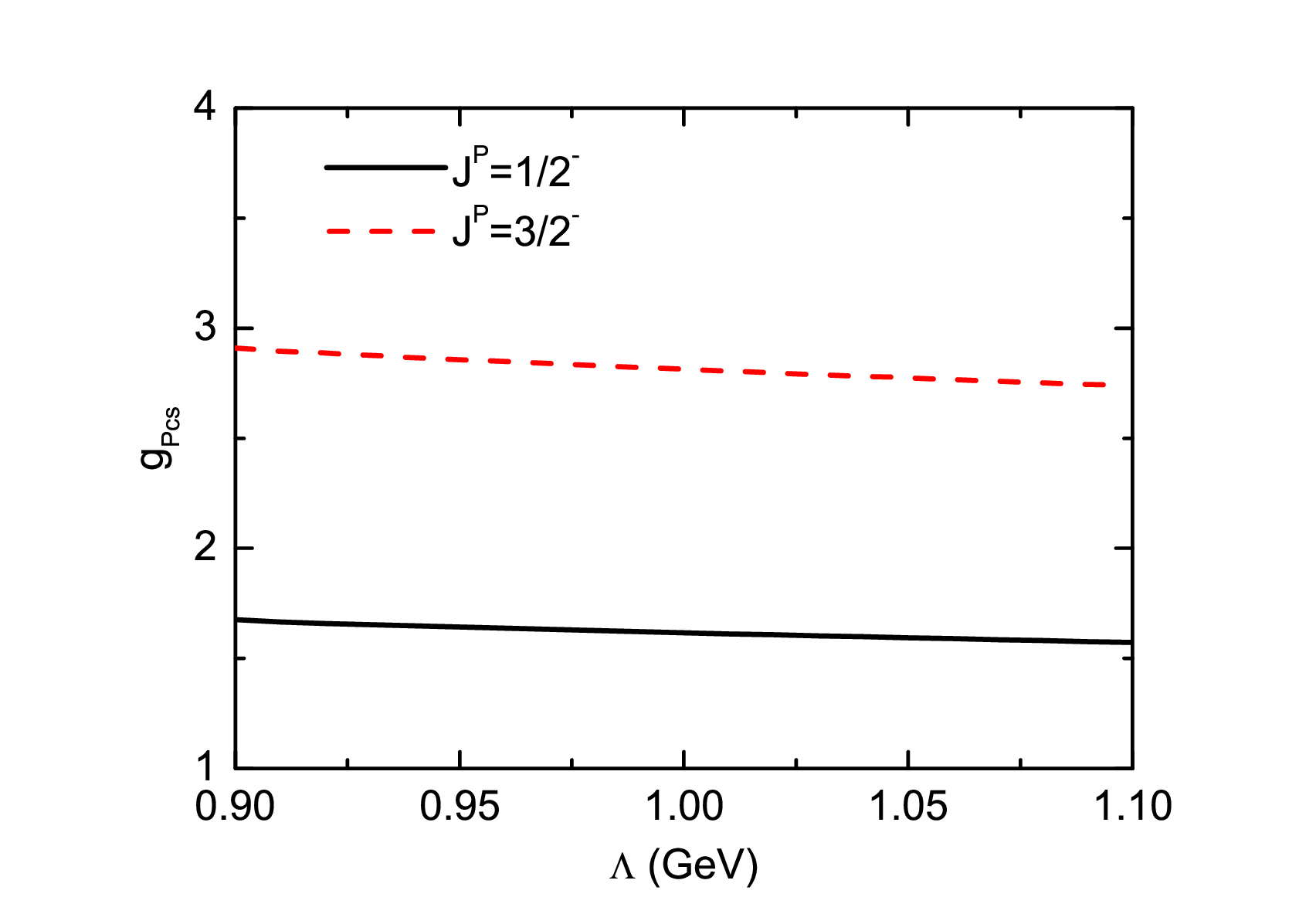}
\caption{(Color online) The coupling constants of the $P_{cs}(4459)$ state with different $J^{P}$
assignments are as a function of the parameter $\Lambda$.}
\label{coupling-constants}
\end{center}
\end{figure}

First, the coupling constants $g_{P_{cs}}$ versus the model parameter $\Lambda$ are computed.
Taking a value of the cutoff $\Lambda=0.9-1.1$ GeV, the corresponding coupling constants are shown in Fig.~\ref{coupling-constants}.
The finding is that they decrease slowly with the increase of the $\Lambda$, and the coupling constants are almost independent of cutoff
$\Lambda$ for the $J^P=1/2^{-}$ and $J^P=3/2^{-}$ cases, where the $P_{cs}(4459)$ is an $S$-wave $\bar{D}^{*}\Xi_c$ molecular state.
This is consistent with the conclusion in Refs.~\cite{Chen:2016qju,Xiao:2016mho,Cleven:2013mka,Dong:2008gb,Dong:2017rmg,Branz:2009yt,Faessler:2007gv}
that for an $S$-wave loosely bound state the effective coupling strength of the bound state to its components is insensitive to its
inner structure.  The results also show that the coupling constant for $J^P=2/3^{-}$ case is bigger than that for the case of $J^P=1/2^{-}$.

According to the discussions in above section and studies in Refs~\cite{Chen:2016qju,Xiao:2016mho,Cleven:2013mka,Dong:2008gb,Dong:2017rmg,Branz:2009yt,Faessler:2007gv},
a typical value of $\Lambda=1.0$ GeV is often employed.  In this work we thus take $\Lambda=1.0$ GeV and the
corresponding coupling constants are listed in Table.~\ref{table1}, which are used to calculate the decay
processes of Fig.~\ref{cc1}.
\begin{table}[h!]
\centering
\caption{ Coupling constants $g_{P_{cs}}$ for different $J^{P}$ states with $\Lambda=1.0$ GeV.
}\label{table1}
\begin{tabular}{cccccccc}
\hline\hline
                               ~~~~~~~~~~~~~ &$J^{P}=1/2^{-}$      ~~~~~~~~~~~~~~~     & $J^{P}=3/2^{-}$~~~~~~~~~~~~~~~          \\
~~~~~~~~~~~~~~~$g_{P_{cs}}$    ~~~~~~~~~~~~~ &1.62                 ~~~~~~~~~~~~~~~     & $2.81         $ ~~~~~~~~~~~~~~~         \\ \hline
\hline
\end{tabular}
\end{table}

\begin{figure}[htbp]
\begin{center}
\includegraphics[bb=10 10 800 400, clip,scale=0.35]{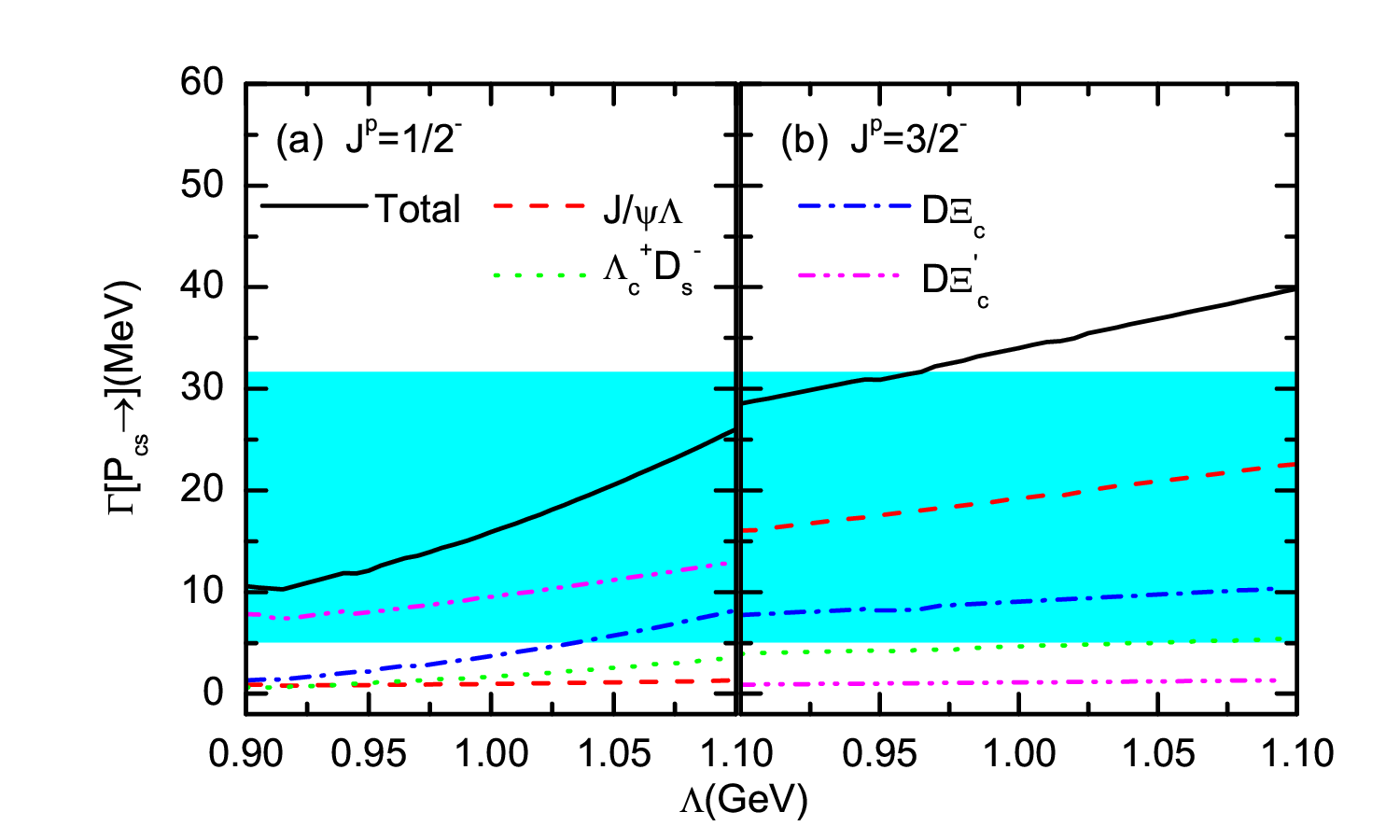}
\caption{(Color online) Partial decay widths of the $P_{cs}\to\Xi_c\bar{D}$ (blue dashed dot
lines), $P_{cs}\to\Lambda{}J/\psi$ (red dash lines), $P_{cs}\to\Lambda_c^{+}D_s^{-}$ (green
dotted lines), $P_{cs}\to\Xi_c^{'}\bar{D}$ (Orange dash dot dot line) ,and the total decay width (black solid lines) with
different spin-parity assignments for the $P_{cs}(4459)$ as a function of the
parameter $\Lambda$. The cyan bands denote the experimental total width~\cite{Aaij:2020gdg}.}
\label{cross-1}
\end{center}
\end{figure}

With the obtained couplings $g_{P_{cs}}$, the total decay width of the $P_{cs}(4459)$ can be calculated straightforwardly.
We show the dependence of the total decay width on the cutoff $\Lambda$ in Fig.~\ref{cross-1}. The cyan bands in these plots
denote the experimental data.  In the present calculation,  we vary $\Lambda$ from 0.9 to 1.1 GeV.  In this cutoff range,
the total decay width increases for the cases of $J^P=1/2^{-}$ and $J^{P}=3/2^{-}$.
For the case of $J^P=3/2^{-}$, the predicted total decay width increases from 28.54 to 39.86 MeV and is slightly bigger than
the experimental total width, which disfavors such a spin-parity assignment for the $P_{cs}(4459)$ in the $\Xi_c\bar{D}^{*}$
molecular picture.  In other words, the decay width of the observed $P_{cs}(4459)$ in $J^P=3/2^{-}$ case cannot be well
reproduced in a pure $\bar{D}^{*}\Xi_c$ molecular state picture.  Following the strategy in Ref.~\cite{Huang:2020taj}, the
$P_{cs}(4459)$ may be a meson-baryon molecule with a lager $\bar{D}^{*}\Xi_c$ component.  This is the same as the results
by He et al.~\cite{Chen:2020kco, Zhu:2021lhd} that the $P_{cs}(4459)$ can be explained as an $S$-wave coupled bound states
with spin-parity $J^P=3/2^{-}$.

The $J^P=1/2^{-}$ assignment is favored by our study.  However, in this case the $\bar{D}^{*}\Xi_c$ molecular state should
be in an $S$-wave.  Hence, only the assignment as an $S$-wave $\Xi_c\bar{D}^{*}$ molecular state with $J^{P}=1/2^{-}$ is
possible for the $P_{cs}(4459)$, based on the total decay width experimentally measured.

We also show the partial decay widths of the $P_{cs}\to\Xi_c\bar{D}$, $P_{cs}\to\Lambda{}J/\psi$,
$P_{cs}\to\Lambda_c^{+}D_s^{-}$, and $P_{cs}\to\Xi_c^{'}\bar{D}$ as a function of the cutoff $\Lambda$ in Fig.~\ref{cross-1}.
The two-body decays are not very sensitive to the cutoff parameter $\Lambda$.  We find that the transition
$P_{cs}\to\Xi_c^{'}\bar{D}$ is the main decay channel for the $J^P=1/2^{-}$ case,
even though the phase space is small compared with the other three channels.  The dominant $P_{cs}\to\Xi_c^{'}\bar{D}$
decay can be easily understood because there is an enhancement due to the mass of the $\Xi_c^{'}\bar{D}$ very close to the
$\Xi_c\bar{D}^{*}$ threshold.  The more important reason is that the long-range $\pi$-meson exchange plays an
indispensable role compared with the other boson exchanges in the interaction between the hadrons.
Only a $\pi$-meson exchange contribution is allowed when studying the nuclear force~\cite{yua:2001ce}.

 \begin{figure}[htbp]
\begin{center}
\includegraphics[bb=50 10 800 315, clip,scale=0.35]{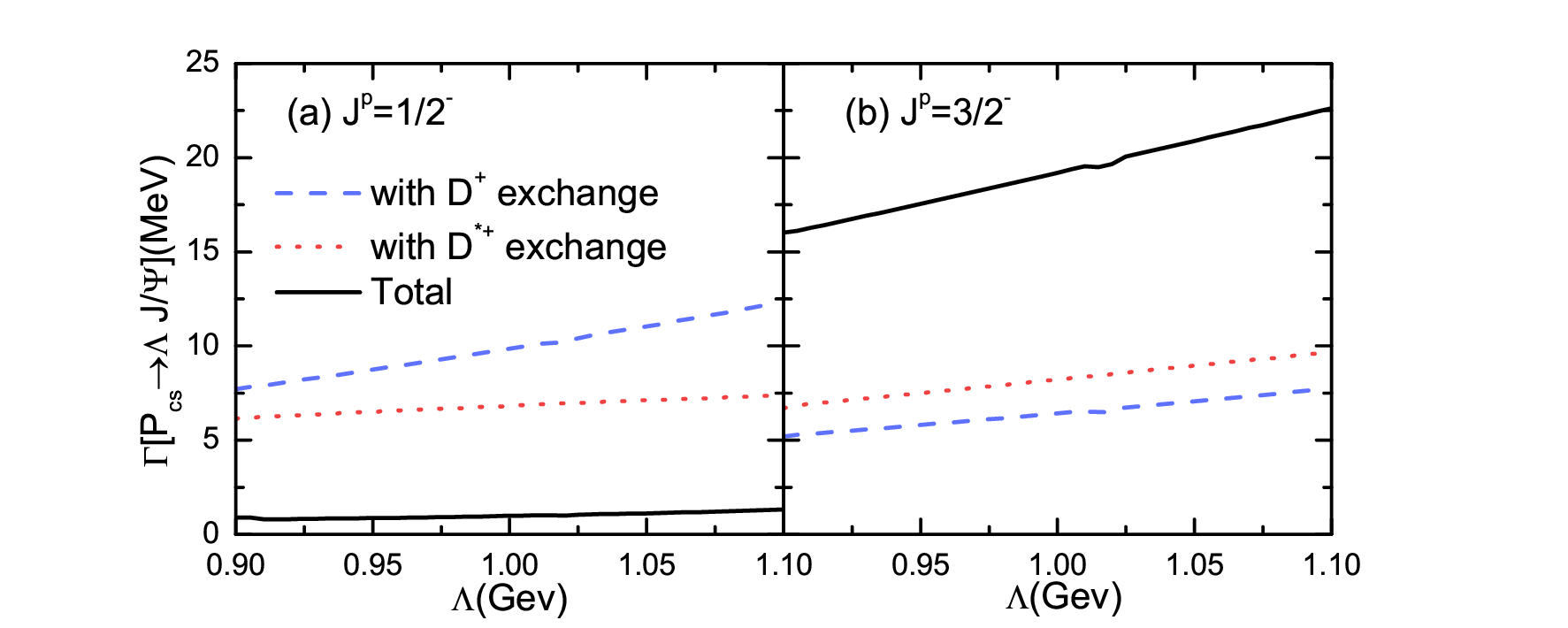}
\caption{(Color online)Individual contributions of the $D^{+}$ and $D^{*+}$ exchange for the $P_{cs}\to\Lambda{}J/\psi$
reaction depending on the parameter $\Lambda$. The red dot and blue dash lines stand for the $D^{+}$ and $D^{*+}$
contributions, respectively.}
\label{cross-2}
\end{center}
\end{figure}

The interference among the individual contributions is sizable, leading to a
total decay width bigger than the partial decay widths in the case of $J^{P}=3/2^{-}$.  It is the exact opposite
of the case of spin parity $J^{P}=1/2^{-}$.  We take the $P_{cs}\to\Lambda{}J/\psi$ as an example and
the results are shown in Fig.~\ref{cross-2}.  This also demonstrates that the $\Gamma{[P_{cs}\to\Lambda{}J/\psi]}$
is the largest for the $J^{P}=3/2^{-}$ case,  while for the $J^{P}=1/2^{-}$ case the transition $P_{cs}\to\Lambda{}J/\psi$
gives minor contributions.

To sum up, the $S$-wave $\bar{D}^{*}\Xi_c$ molecular with $J^{P}=1/2^{-}$ assignment for $P_{cs}(4459)$ is supported
by our study, while the $P_{cs}(4459)$ in spin-parity $J^P=3/2^{-}$ case may be explained
as an $S$-wave coupled bound state with lager $\Xi_c\bar{D}^{*}$ component. It is worth noting that the $P_{cs}(4459)$
can also be considered as a compact pentaquark state~\cite{Wang:2020eep,Azizi:2021utt}.
Theoretical investigations on other decay modes and further experimental information on its spin parities and
partial decay widths will be very helpful to understand the nature of the $P_{cs}(4459)$.

\section{SUMMARY}
We studied the strong decays of the newly observed $P_{cs}(4459)$ baryon into $J/\psi{}\Lambda$, $D_s^{-}\Lambda_c^{+}$, and
$\bar{D}\Xi_c^{(')}$ with different spin-parity assignments, assuming that it is a $\bar{D}^{*}\Xi_c$
molecular structure. With the coupling constants between the $P_{cs}(4459)$ and its components determined by the composition
condition, we calculated the partial decay widths into $J/\psi{}\Lambda$, $D_s^{-}\Lambda_c^{+}$, and $\bar{D}\Xi_c^{(')}$ final
states through triangle diagrams in an effective Lagrangian approach. In such a picture, the decays $P_{cs}\to\Xi_c\bar{D}$,
$P_{cs}\to\Lambda{}J/\psi$, $P_{cs}\to\Lambda_c^{+}D_s^{-}$, and $P_{cs}\to\Xi_c^{'}\bar{D}$ occur by exchanging $\pi$, $D^{(*)}$,
and $K^{(*)}$ mesons.  We found that the total decay width can be reproduced with the assumption that the $P_{cs}(4459)$
is an $S$-wave $\bar{D}^{*}\Xi_c$ molecular with $J^{P}=1/2^{-}$ assignment, while the $P_{cs}(4459)$ in spin-parity
$J^P=3/2^{-}$ case maybe explained as an $S$-wave coupled bound state with lager $\Xi_c\bar{D}^{*}$ component.
Our study shows that the experimental measurement of the spin-parity and the $\bar{D}\Xi_c^{'}$ decay width of the $P_{cs}(4459)$
will be able to tell whether it is a pure $\Xi_c\bar{D}^{*}$ molecular state, coupled bound state with
lager $\Xi_c\bar{D}^{*}$ component, or a compact pentaquark state.

\section*{Acknowledgments}
This work was supported by the Science and Technology
Research Program of Chongqing Municipal Education Commission (Grant No. KJQN201800510), the Opened Fund
of the State Key Laboratory on Integrated Optoelectronics
(GrantNo. IOSKL2017KF19). Yin Huang want to thanks
the support from the Development and Exchange Platform for
the Theoretic Physics of Southwest Jiaotong University under Grants No.11947404 and No.12047576,
the Fundamental Research Funds for the Central Universities(Grant No.
2682020CX70), and the National Natural Science Foundation
of China under Grant No.12005177.


\begin{thebibliography}{90}
%\cite{Zyla:2020zbs}
\bibitem{Zyla:2020zbs}
P.~A.~Zyla \textit{et al.} [Particle Data Group],
%``Review of Particle Physics,''
PTEP \textbf{2020}, 083C01 (2020).
%1495 citations counted in INSPIRE as of 04 Jun 2021


%\cite{Aaij:2019vzc}
\bibitem{Aaij:2019vzc}
R.~Aaij \textit{et al.} [LHCb],
%``Observation of a narrow pentaquark state, $P_c(4312)^+$, and of two-peak structure of the $P_c(4450)^+$,''
Phys. Rev. Lett. \textbf{122}, 222001 (2019).
%306 citations counted in INSPIRE as of 04 Jun 2021


%\cite{Aaij:2020gdg}
\bibitem{Aaij:2020gdg}
R.~Aaij \textit{et al.} [LHCb],
%``Evidence of a $J/\psi \varLambda$ structure and observation of excited $\varXi^-$ states in the $\varXi_b^-\to J/\psi \varLambda K^-$ decay,''
[arXiv:2012.10380 [hep-ex]].
%26 citations counted in INSPIRE as of 04 Jun 2021



%\cite{Wu:2010vk}
\bibitem{Wu:2010vk}
J.~J.~Wu, R.~Molina, E.~Oset and B.~S.~Zou,
%``Dynamically generated $N^{*}$ and $\Lambda^*$ resonances in the hidden charm sector around 4.3 GeV,''
Phys. Rev. C \textbf{84}, 015202 (2011).
%193 citations counted in INSPIRE as of 23 Jul 2021


%\cite{Wang:2019nvm}
\bibitem{Wang:2019nvm}
B.~Wang, L.~Meng and S.~L.~Zhu,
%``Spectrum of the strange hidden charm molecular pentaquarks in chiral effective field theory,''
Phys. Rev. D \textbf{101}, 034018 (2020)
doi:10.1103/PhysRevD.101.034018
[arXiv:1912.12592 [hep-ph]].
%25 citations counted in INSPIRE as of 23 Jul 2021


%\cite{Anisovich:2015zqa}
\bibitem{Anisovich:2015zqa}
V.~V.~Anisovich, M.~A.~Matveev, J.~Nyiri, A.~V.~Sarantsev and A.~N.~Semenova,
%``Nonstrange and strange pentaquarks with hidden charm,''
Int. J. Mod. Phys. A \textbf{30}, 1550190 (2015).
%44 citations counted in INSPIRE as of 23 Jul 2021

%\cite{Wang:2015wsa}
\bibitem{Wang:2015wsa}
Z.~G.~Wang,
%``Analysis of the ${\frac{1}{2}}^{\pm }$ pentaquark states in the diquark\textendash{}diquark\textendash{}antiquark model with QCD sum rules,''
Eur. Phys. J. C \textbf{76},142 (2016).
%37 citations counted in INSPIRE as of 23 Jul 2021

%\cite{Feijoo:2015kts}
\bibitem{Feijoo:2015kts}
A.~Feijoo, V.~K.~Magas, A.~Ramos and E.~Oset,
%``A hidden-charm $S=-1$ pentaquark from the decay of $\varLambda _b$ into $J/\psi, \eta \varLambda$ states,''
Eur. Phys. J. C \textbf{76}, 446 (2016).
%45 citations counted in INSPIRE as of 23 Jul 2021


%\cite{Lu:2016roh}
\bibitem{Lu:2016roh}
J.~X.~Lu, E.~Wang, J.~J.~Xie, L.~S.~Geng and E.~Oset,
%``The $\Lambda_{b}\rightarrow J/\psi K^{0}\Lambda$ reaction and a hidden-charm pentaquark state with strangeness,''
Phys. Rev. D \textbf{93}, 094009 (2016).
%51 citations counted in INSPIRE as of 23 Jul 2021

%\cite{Chen:2016ryt}
\bibitem{Chen:2016ryt}
R.~Chen, J.~He and X.~Liu,
%``Possible strange hidden-charm pentaquarks from $\Sigma_c^{(*)}\bar{D}_s^*$ and $\Xi^{(',*)}_c\bar{D}^*$ interactions,''
Chin. Phys. C \textbf{41}, 103105 (2017).
%30 citations counted in INSPIRE as of 23 Jul 2021

%\cite{Xiao:2019gjd}
\bibitem{Xiao:2019gjd}
C.~W.~Xiao, J.~Nieves and E.~Oset,
%``Prediction of hidden charm strange molecular baryon states with heavy quark spin symmetry,''
Phys. Lett. B \textbf{799}, 135051 (2019).
%34 citations counted in INSPIRE as of 23 Jul 2021

%\cite{Chen:2015sxa}
\bibitem{Chen:2015sxa}
H.~X.~Chen, L.~S.~Geng, W.~H.~Liang, E.~Oset, E.~Wang and J.~J.~Xie,
%``Looking for a hidden-charm pentaquark state with strangeness S=\ensuremath{-}1 from \ensuremath{\Xi}$_b^?$ decay into J/\ensuremath{\psi}K$^?$\ensuremath{\Lambda},''
Phys. Rev. C \textbf{93}, 065203 (2016).
%61 citations counted in INSPIRE as of 23 Jul 2021













%\cite{Wang:2020eep}
\bibitem{Wang:2020eep}
Z.~G.~Wang,
%``Analysis of the $P_{cs}(4459)$ as the hidden-charm pentaquark state with QCD sum rules,''
Int. J. Mod. Phys. A \textbf{36}, 2150071 (2021).
%15 citations counted in INSPIRE as of 22 Jul 2021


%\cite{Azizi:2021utt}
\bibitem{Azizi:2021utt}
K.~Azizi, Y.~Sarac and H.~Sundu,
%``Investigation of $P_{cs}(4459)^0$ pentaquark via its strong decay to $\Lambda J/\Psi$,''
Phys. Rev. D \textbf{103}, 094033 (2021).
%5 citations counted in INSPIRE as of 22 Jul 2021


%\cite{Peng:2020hql}
\bibitem{Peng:2020hql}
F.~Z.~Peng, M.~J.~Yan, M.~S\'anchez S\'anchez and M.~P.~Valderrama,
%``The $P_{cs}(4459)$ pentaquark from a combined effective field theory and phenomenological perspectives,''
[arXiv:2011.01915 [hep-ph]].
%22 citations counted in INSPIRE as of 23 Jul 2021


%\cite{Chen:2020uif}
\bibitem{Chen:2020uif}
H.~X.~Chen, W.~Chen, X.~Liu and X.~H.~Liu,
%``Establishing the first hidden-charm pentaquark with strangeness,''
Eur. Phys. J. C \textbf{81}, 409 (2021).
%21 citations counted in INSPIRE as of 23 Jul 2021


%\cite{Chen:2020kco}
\bibitem{Chen:2020kco}
R.~Chen,
%``Can the newly reported $P_{cs}(4459)$ be a strange hidden-charm $\Xi_c\bar D^*$ molecular pentaquark?,''
Phys. Rev. D \textbf{103}, 054007 (2021).
%14 citations counted in INSPIRE as of 23 Jul 2021


%\cite{Zhu:2021lhd}
\bibitem{Zhu:2021lhd}
J.~T.~Zhu, L.~Q.~Song and J.~He,
%``$P_{cs}(4459)$ and other possible molecular states from $\Xi_{c}^{(*)}\bar{D}^{(*)}$ and $\Xi'_c\bar{D}^{(*)}$ interactions,''
Phys. Rev. D \textbf{103}, 074007 (2021).
%8 citations counted in INSPIRE as of 23 Jul 2021


%\cite{Lu:2021irg}
\bibitem{Lu:2021irg}
J.~X.~Lu, M.~Z.~Liu, R.~X.~Shi and L.~S.~Geng,
%``Understanding $P_{cs}(4459)$ as a hadronic molecule in the $\Xi_b^-\to J/\psi \Lambda K^-$ decay,''
[arXiv:2104.10303 [hep-ph]].
%4 citations counted in INSPIRE as of 23 Jul 2021


%\cite{Xiao:2019mvs}
\bibitem{Xiao:2019mvs}
C.~J.~Xiao, Y.~Huang, Y.~B.~Dong, L.~S.~Geng and D.~Y.~Chen,
%``Exploring the molecular scenario of Pc(4312) , Pc(4440) , and Pc(4457),''
Phys. Rev. D \textbf{100}, 014022 (2019).
%85 citations counted in INSPIRE as of 27 Jul 2021



%\cite{Dong:2008gb}
\bibitem{Dong:2008gb}
  Y.~Dong, A.~Faessler, T.~Gutsche and V.~E.~Lyubovitskij,
%  ``Estimate for the $X(3872) \to \gamma J/\psi$ decay width,''
  Phys.\ Rev.\ D {\bf 77}, 094013 (2008).
  %%CITATION = doi:10.1103/PhysRevD.77.094013;%%
  %109 citations counted in INSPIRE as of 04 Dec 2017

%\cite{Dong:2017rmg}
\bibitem{Dong:2017rmg}
  Y.~Dong, A.~Faessler, T.~Gutsche, Q.~Lu  and V.~E.~Lyubovitskij,
%  ``Selected strong decays of $\eta(2225)$ and $\phi(2170)$ as $\Lambda \bar\Lambda$ bound states,''
  Phys.\ Rev.\ D {\bf 96}, no. 7, 074027 (2017).
  %%CITATION = doi:10.1103/PhysRevD.96.074027;%%



  %\cite{Xiao:2016mho}
\bibitem{Xiao:2016mho}
  C.~J.~Xiao and D.~Y.~Chen,
  %``Possible $B^{(\ast)} \bar{K}$ hadronic molecule state,''
  Eur.\ Phys.\ J.\ A {\bf 53}, no. 6, 127 (2017).
  %%CITATION = doi:10.1140/epja/i2017-12310-x;%%
  %26 citations counted in INSPIRE as of 26 Dec 2017

%\cite{Cleven:2013mka}
\bibitem{Cleven:2013mka}
  M.~Cleven, Q.~Wang, F.~K.~Guo, C.~Hanhart, Meissner.Ulf-G and Q.~Zhao,
  %``$Y(4260)$ as the first $S$-wave open charm vector molecular state?,''
  Phys.\ Rev.\ D {\bf 90}, no. 7, 074039 (2014).
  %%CITATION = doi:10.1103/PhysRevD.90.074039;%%
  %44 citations counted in INSPIRE as of 10 Jan 2018


%\cite{Chen:2016qju}
\bibitem{Chen:2016qju}
  H.~X.~Chen, W.~Chen, X.~Liu and S.~L.~Zhu,
  %``The hidden-charm pentaquark and tetraquark states,''
  Phys.\ Rept.\  {\bf 639}, 1 (2016).
  %%CITATION = doi:10.1016/j.physrep.2016.05.004;%%
  %221 citations counted in INSPIRE as of 26 Dec 2017

  %\cite{Branz:2009yt}
\bibitem{Branz:2009yt}
  T.~Branz, T.~Gutsche and V.~E.~Lyubovitskij,
  %``Hadronic molecule structure of the Y(3940) and Y(4140),''
  Phys.\ Rev.\ D {\bf 80}, 054019 (2009).
  %%CITATION = doi:10.1103/PhysRevD.80.054019;%%
  %158 citations counted in INSPIRE as of 26 Dec 2017

  %\cite{Faessler:2007gv}
\bibitem{Faessler:2007gv}
  A.~Faessler, T.~Gutsche, V.~E.~Lyubovitskij and Y.~L.~Ma,
  %``Strong and radiative decays of the D(s0)*(2317) meson in the DK-molecule picture,''
  Phys.\ Rev.\ D {\bf 76}, 014005 (2007).
  %%CITATION = doi:10.1103/PhysRevD.76.014005;%%
  %123 citations counted in INSPIRE as of 26 Dec 2017

%\cite{Dong:2009tg}
\bibitem{Dong:2009tg}
  Y.~Dong, A.~Faessler, T.~Gutsche and V.~E.~Lyubovitskij,
%  ``Strong two-body decays of the Lambda(c)(2940)+ in a hadronic molecule picture,''
  Phys.\ Rev.\ D {\bf 81}, 014006 (2010).
  %%CITATION = doi:10.1103/PhysRevD.81.014006;%%
  %41 citations counted in INSPIRE as of 23 Nov 2017

%\cite{Weinberg:1962hj}
\bibitem{Weinberg:1962hj}
  S.~Weinberg,
%  ``Elementary particle theory of composite particles,''
  Phys.\ Rev.\  {\bf 130}, 776 (1963).
  %%CITATION = doi:10.1103/PhysRev.130.776;%%
  %444 citations counted in INSPIRE as of 20 Nov 2017

%\cite{Salam:1962ap}
\bibitem{Salam:1962ap}
  A.~Salam,
%  ``Lagrangian theory of composite particles,''
  Nuovo Cim.\  {\bf 25}, 224 (1962).
  %%CITATION = doi:10.1007/BF02733330;%%
  %215 citations counted in INSPIRE as of 20 Nov 2017

%\cite{Oller:2017alp}
\bibitem{Oller:2017alp}
J.~A.~Oller,
%``New results from a number operator interpretation of the compositeness of bound and resonant states,''
Annals Phys. \textbf{396}, 429-458 (2018).
%21 citations counted in INSPIRE as of 18 Sep 2021


%\cite{Guo:2015daa}
\bibitem{Guo:2015daa}
Z.~H.~Guo and J.~A.~Oller,
%``Probabilistic interpretation of compositeness relation for resonances,''
Phys. Rev. D \textbf{93},096001 (2016).
%84 citations counted in INSPIRE as of 18 Sep 2021


%\cite{Chen:2021tip}
\bibitem{Chen:2021tip}
R.~Chen,
%``Strong decays of the newly $P_{cs}(4459)$ as a strange hidden-charm $\Xi _c{\bar{D}}^*$ molecule,''
Eur. Phys. J. C \textbf{81}, 122 (2021).
%7 citations counted in INSPIRE as of 22 Jul 2021


%\cite{Liu:2001ce}
\bibitem{Liu:2001ce}
W.~Liu, C.~M.~Ko and Z.~W.~Lin,
%``Cross-section for charmonium absorption by nucleons,''
Phys. Rev. C \textbf{65}, 015203 (2002).
%72 citations counted in INSPIRE as of 22 Jul 2021


%\cite{Huang:2020taj}
\bibitem{Huang:2020taj}
Y.~Huang and L.~Geng,
%``Strong decays of the $\Xi(1620)$ as a $\Lambda\bar{K}$ and $\Sigma\bar{K}$ molecule,''
Eur. Phys. J. C \textbf{80}, 837 (2020).
%1 citations counted in INSPIRE as of 27 Jul 2021


%\cite{yua:2001ce}
\bibitem{yua:2001ce}
H. Yukawa, Proc. Phys. Math. Soc. Jap. 17, 48 (1935), [Prog. Theor. Phys. Suppl.1,1(1935)].



\end{thebibliography}
\end{document}